\documentclass{emulateapj}
\usepackage{apjfonts,graphicx,ifpdf}
\usepackage{color}

\def\etal{{\it et~al.}}

\newcommand{\water}{\mbox{H$_2$O}}

\newcommand{\mum}{$\mu$m}
\newcommand{\muas}{$\mu$as}
\newcommand{\muasyr}{$\mu$as yr$^{-1}$}

\newcommand{\kms}{km~s$^{-1}$}

\shorttitle{Water Masers in the Andromeda Galaxy}
\shortauthors{Darling}

\begin{document}

\title{Water Masers in the Andromeda Galaxy:  The First Step Toward Proper Motion}

\author{Jeremy Darling\altaffilmark{1}}
\altaffiltext{1}{Center for Astrophysics and Space Astronomy,
Department of Astrophysical and Planetary Sciences,
University of Colorado, 389 UCB, Boulder, CO 80309-0389; 
jdarling@origins.colorado.edu}

\begin{abstract}
We have detected and confirmed five water maser complexes in the 
Andromeda Galaxy (M31) using the Green Bank Telescope.  These masers will provide the high brightness
temperature point sources needed for proper motion studies of M31, 
enabling measurement of its full three-dimensional velocity vector and its geometric 
distance via proper rotation.  The motion
of M31 is the keystone of Local Group dynamics and a gateway to the 
dark matter profiles of galaxies in general.
Our survey for water masers 
selected 206 luminous compact 24~\mum-emitting 
regions in M31 and was sensitive enough to detect any maser useful for 
$\sim$10~\muasyr\ astrometry.
The newly discovered masers span the isotropic luminosity
range 0.3--1.9 $\times 10^{-3}\,L_\odot$ in 
single spectral components and are analogous to luminous Galactic masers.
The masers %sample most of the disk rotation of M31 and 
are distributed around the molecular ring, including locations close to the major
and minor axes, which is nearly ideal for proper motion studies.  
We find no correlation between 24~\mum\ luminosity and 
water maser luminosity, suggesting that while water masers arise in 
star-forming regions, the nonlinear amplification pathways and beamed
nature of the water masers means that they are not predictable based on IR 
luminosity alone.  This
suggests that there are additional bright masers to be found in M31.
%, which can provide an astrometry framework for proper motion studies.
We predict that the geometric distance and systemic proper motion of M31 can
be measured in 2--3 years with current facilities.  A ``moving
cluster'' observation of diverging masers as M31 approaches the Galaxy
may be possible in the long term.
\end{abstract}


\section{Introduction}

The dominant galaxies in the Local Group, 
the Milky Way and Andromeda (M31), are likely to collide 
in the next 5--10 Gyr \citep{loeb05}.  
%These $\sim L^*$ galaxies are likely to become a 
%ULIRG during merging and may even produce an OH megamaser.  
In the meantime, 
however, we would like to know the proper (transverse) motion of M31 with 
respect to the Milky Way. %(the galaxies have a relative velocity of 
%$-$117~km~s$^{-1}$ and a separation of $\sim780$~kpc {\red ref}).  
\citet{peebles01} show that measurements of the transverse motions of Local 
Group galaxies will critically test the fundamental assumption that luminosities 
trace the mass of galaxies.  The proper motion of M31 is key to our understanding of 
the Local Group's future (and past) dynamical evolution as well as the density 
profiles and distribution of dark matter halos \citep{loeb05,reid09}. 

%\noindent
\citet{loeb05} suggest that GAIA, SIM, or the Square Kilometer Array 
will provide the $\sim$10--20~$\mu$as~yr$^{-1}$ astrometry required 
for a proper motion determination of M31 \citep[e.g.,][]{vandermarel08}.
But Galactic analog masers in M31 observed with sensitive Very Long Baseline 
Interferometers (VLBI) over a time baseline of a few years would be  
adequate.  Indeed, 22~GHz water (H$_2$O) masers
have been discovered in M31's
satellite galaxies M33 \citep{churchwell77} and IC~10 \citep{henkel86}, and 
\citet{brunthaler05,brunthaler07} have 
measured their proper motions, yielding their 3-dimensional
velocities with respect to the Milky Way.  \citet{brunthaler05} also measured the proper rotation of 
M33, which provides a geometric determination of its distance via ``rotational parallax.''
Loeb \etal\ (2005) show that a 
backwards-time-evolution model has M33 colliding with M31 in the past.
Since M33 does not bear the morphological signatures of such an interaction
(but does show star formation possibly triggered by a close pass), 
one must assume that M31 was elsewhere at the time and thus has a 
non-negligible tangential motion, of order 100~\kms\  
\citep[][but see \citet{vandermarel08}]{loeb05,brunthaler07}.
%\citep[][but see van der Marel \& Guhathakurta 2008]{loeb05,brunthaler07}.

%\noindent
Detecting masers in M31 has historically been problematic \citep{sullivan73}.
\citet{greenhill95} and \citet{imai01} conducted pointed
H$_2$O maser surveys toward \ion{H}{2} regions in M31, reaching 1$\sigma$ rms noise levels of 
29 and 70~mJy, respectively.  
Claussen \& Beasley (private communication) conducted a survey for H$_2$O 
masers in the nuclear region and across most of the molecular ring
with the Very Large Array, 
making no detections and reaching an rms noise of 30 mJy per beam.
\citet{sjouwerman10} have detected the first maser of any type in M31, 
a Class II methanol (CH$_3$OH) maser at 6.7~GHz, but no
%Follow-up observations for 
associated water masers were detected.
% at the location of their discovery were not successful.  
The water maser non-detections to date are all consistent with 
the aggregate star formation rate of M31, and these surveys 
were not sufficiently sensitive to have detected a typical Galactic maser 
at the distance of M31 \citep{brunthaler06}.  
The lack of Galactic analog water maser detections in M31 has simply
been a problem of sensitivity and of locating likely sites of maser activity, both of which
now have remedies.

\begin{figure*}[ht!]
\epsscale{1.18}
\plotone{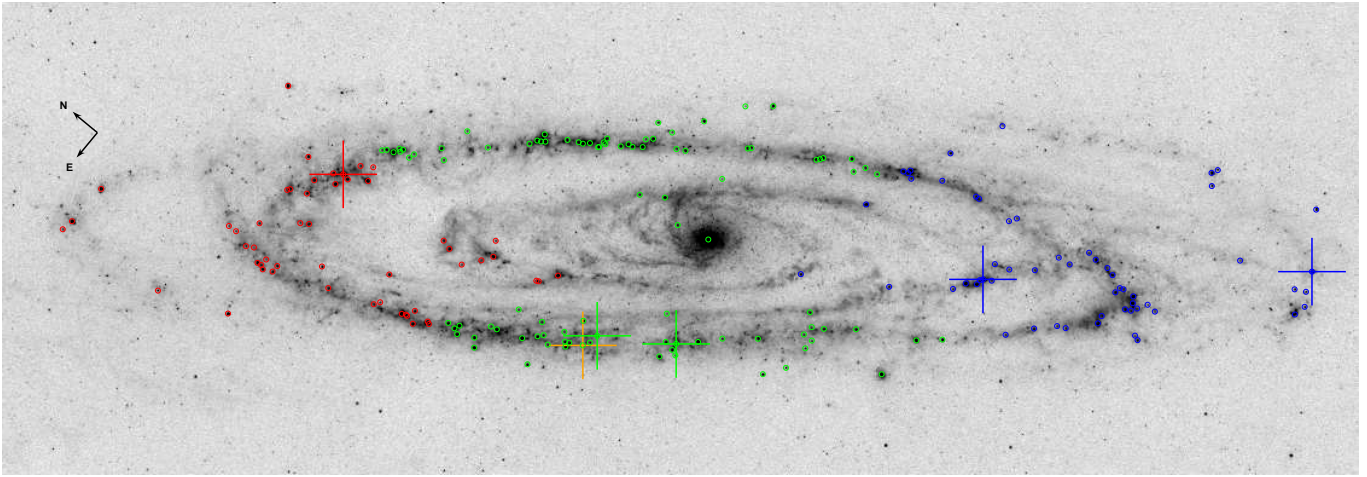}
\caption{
Spitzer 24~$\mu$m map of M31 \citep{gordon08}.
The circles indicate the 206 pointing centers 
%observed with the 33'' FWHM beam of the GBT K-band dual feed system 
(the circles are to scale, showing the 33\arcsec\ FWHM beam).  
Crosses mark the detected masers and are enlarged for clarity.  The orange
cross indicates the 6.7~GHz methanol maser detected by \citet{sjouwerman10}.
%we confirm the absence of a water maser in this region.
Colors indicate the systemic (green), red, and blue spectrometer
tuning centers at $-$300, $-$100, and $-$500~\kms, respectively.
% Remove all unobserved sources.
%.  Indicate detections. Add coordinates and compass.
 %\scriptsize
 }\label{fig:M31map}
\end{figure*}

Galactic analog H$_2$O masers associated with star formation
are detectable in M31 in reasonable 
integration time using the Green Bank Telescope\footnote{The National
  Radio Astronomy Observatory is a facility of the National Science
  Foundation operated under cooperative agreement by Associated
  Universities, Inc.}  (GBT); 
they already have been detected in significantly more distant
galaxies (Darling, Brogan, \& Johnson 2008).  
But a full survey of the star-forming regions of M31 with the sensitivity to detect 
Galactic analog water masers is prohibitive because the galaxy is so large on the sky.
We thus combined an order of magnitude sensitivity improvement compared to
the best previous survey with a physically motivated pointing guide to the most
likely sites of water maser activity.  In this Letter, we describe the source selection,
%in \S \ref{sec:selection}, 
the observations and data reduction,
% in \S \ref{sec:obs}, 
and the newly discovered water masers.
% in \S \ref{sec:results}, 
We also discuss the prospects for using these masers for 
proper motion studies of M31.
% in \S \ref{sec:discussion}.

Throughout this Letter, we use heliocentric velocities (optical definition)
and assume the following about M31:
a systemic velocity of $-300$~\kms\ \citep{devaucouleurs91},
a distance of 780~kpc \citep{mcconnachie05}, 
a major axis position angle of $38^\circ$ (E of N), 
a constant inclination of $77^\circ$, 
and a central coordinates 00:42:44.3, +41:16:09 (J2000).

\section{Sample Selection}\label{sec:selection}
 
A Spitzer 24~$\mu$m map of M31 \citep{gordon08}
guided the GBT observations in two important ways:  source selection and completeness.  
The association of water masers with (ultra)compact \ion{H}{2} regions is well known, but
%recent mapping of extragalactic water masers has demonstrated that
%peaks in 
the molecular gas as traced by CO emission does not necessarily indicate these regions.
%\citep[e.g.,][]{brogan10} {\red [Galactic ref]}.  
Likewise, H$\alpha$-selected \ion{H}{2} regions may select {\it against} 
dust-enshrouded ultra-compact \ion{H}{2} regions. % because observable
%H$\alpha$ emission selects against dust extinction. % \citep{brogan10}.
Compact 24~$\mu$m emission, however, is a good indicator of likely
H$_2$O masers, and the rough relationship between far-IR emission and 
water maser luminosity is well-known %\citep{darling08} 
\citep[e.g.,][]{jaffe81,castangia08}.
We therefore selected compact (unresolved at the 23.5~pc resolution of the Spitzer map)  24~$\mu$m sources 
associated with the dusty molecular --- and presumably star-forming --- regions in M31.  We 
constructed a catalog of 206 of these objects working from the brightest down to the point where most 
of the 24~$\mu$m emission becomes extended, about 4~MJy~steradian$^{-1}$ at peak specific 
intensity.
Although this pointed survey does not include all star formation in
M31, it does include a large fraction of the likely sites for  strong H$_2$O maser activity, and by working down 
in luminosity through the catalog of compact 24~$\mu$m sources, we have surveyed a large fraction of the 
total ongoing star formation in the galaxy (e.g., Tabatabaei \& Berkhuijsen 2010) and thus
 most of the likely H$_2$O-bearing regions (\S \ref{sec:discussion}).
Figure \ref{fig:M31map} shows the 24~$\mu$m map of M31 with the
pointing centers and primary beam size
for our GBT water maser survey.

%We have conducted the first systematic, sensitive water maser survey of M31 that is likely
%to detect maser activity consistent with the galaxy's current star formation rate.  

%{\rotate
\begin{deluxetable*}{|ccc|ccc|ccc|ccc|}
\tabletypesize{\scriptsize} 
\tablecaption{Water Maser Survey Sample\label{tab:obs}} 
\tablewidth{0pt} 
\tablehead{
\colhead{Source} & \colhead{$V_{\rm Obs}$} & \colhead{rms} & 
\colhead{Source} & \colhead{$V_{\rm Obs}$} & \colhead{rms} & 
\colhead{Source} & \colhead{$V_{\rm Obs}$} & \colhead{rms} & 
\colhead{Source} & \colhead{$V_{\rm Obs}$} & \colhead{rms} \\
\colhead{J2000} & \colhead{km s$^{-1}$} & \colhead{mJy} & 
\colhead{J2000} & \colhead{km s$^{-1}$} & \colhead{mJy} & 
\colhead{J2000} & \colhead{km s$^{-1}$} & \colhead{mJy} & 
\colhead{J2000} & \colhead{km s$^{-1}$} & \colhead{mJy}}
\startdata 
003849.2+402551.7 & $-$500 & 3.2 & 

003906.7+403704.5 & $-$500 & 3.5 & 
003910.2+403725.6 & $-$500 & 3.2 & 
003916.1+403629.5 & $-$500 & 3.0 \\
{\bf 003918.9+402158.4} & $-500$ & 1.8 & 
003930.2+402106.4 & $-$500 & 3.7 &   
003933.2+402215.6 & $-$500 & 3.2 &  
003937.5+402011.5 & $-$500 & 3.4 \\
003939.8+402856.3 & $-$500 & 3.1 & 
003944.5+402030.4 & $-$500 & 3.3 & 

004004.7+405840.9 & $-$500 & 3.2 & 
004030.9+404230.0 & $-$500 & 3.0 \\
004031.2+403952.0 & $-$500 & 2.9 &  
004031.7+404127.0 & $-$500 & 3.3 &  
004032.6+403856.1 & $-$500 & 3.3 &  
004033.3+403352.1 & $-$500 & 3.1 \\
004033.8+403246.6 & $-$500 & 3.1 & 
004034.7+403541.2 & $-$500 & 2.8 &  
004035.1+403701.1 & $-$500 & 4.0 & 
004035.8+403724.6 & $-$500 & 3.6 \\ 
{\it 004036.1+410117.5} & $-$500 & 2.1 &
004038.0+403514.9 & $-$500 & 2.9 & 
004038.8+403431.0 & $-$500 & 3.4 &  
004039.4+403730.5 & $-$500 & 3.4 \\
{\it 004041.6+405105.0} & $-$500 & 2.0 & 
004042.1+403454.5 & $-$500 & 3.4 &  
004043.3+404321.9 & $-$500 & 3.1 &  
004043.6+403530.5 & $-$500 & 3.3 \\ 
004044.2+404446.4 & $-$500 & 3.0 & 
{\it 004045.7+405134.5} & $-$500 & 1.9 &  
004046.4+405541.9 & $-$500 & 3.0 & 
004046.5+405606.4 & $-$500 & 2.8 \\
004051.6+410006.5 & $-$500 & 2.9 &  
004051.9+403249.7 & $-$500 & 3.4 &  
004053.0+403218.0 & $-$500 & 3.8 &  
004055.1+403703.2 & $-$500 & 3.8 \\
{\it 004058.6+404558.0} & $-$500 & 2.1 & 
004058.6+410332.3 & $-$500 & 3.0 & 
004059.8+403652.4 & $-$500 & 4.8 & 

004100.6+410334.0 & $-$500 & 3.7 \\
{\it 004101.6+410405.8} & $-$500 & 2.9 & 
{\it 004102.0+410254.9} & $-$500 & 1.7 & 
004107.6+404812.5 & $-$500 & 3.5 & 
004110.4+404949.5 & $-$500 & 3.1 \\ 
004112.5+410609.7 & $-$300 & 3.4 & 
{\it 004113.7+403918.6} & $-$500 & 2.2 &
{\it 004113.9+410736.1} & $-$300 & 2.2 & 
004114.8+410923.7 & $-$300 & 3.8 \\
004115.9+404011.6 & $-$500 & 3.1 & 
004119.1+404857.4 & $-$500 & 3.6 &  
004119.5+411948.8 & $-$300 & 3.9 & 
004120.0+410821.5 & $-$300 & 3.0 \\
{\bf 004121.7+404947.7} & $-500$ & 2.4 & 
004123.2+405000.6 & $-$500 & 3.4 &  
004124.8+411154.6 & $-$300 & 3.4 & 
004125.4+404200.4 & $-$500 & 3.7 \\
004126.1+404959.1 & $-$500 & 3.7 & 
004126.5+411206.9 & $-$300 & 3.0 & 
004128.1+411222.6 & $-$300 & 3.6 & 
{\it 004129.8+405059.5} & $-$500 & 2.2 \\
004129.8+412211.1 & $-$300 & 3.1 & 
004130.3+410501.7 & $-$500 & 3.1 & 
004137.0+405142.5 & $-$500 & 3.3 &   
004138.6+404401.2 & $-$500 & 3.4 \\
004146.7+411846.6 & $-$300 & 3.3 & 
004148.2+411903.8 & $-$300 & 3.6 & 
004151.9+412442.1 & $-$300 & 3.3 & 
{\it 004159.4+405720.8} & $-$500 & 2.1 \\

{\it 004203.9+404907.1} & $-$300 & 3.7 & 
004208.8+412639.9 & $-$300 & 3.2 & 
004209.5+412832.3 & $-$300 & 3.3 & 
004211.6+411909.4 & $-$300 & 2.7 \\
004212.3+412415.7 & $-$300 & 2.9 & 
004213.8+405117.7 & $-$300 & 3.9 & 
004214.8+412508.9 & $-$300 & 2.8 &  
004218.7+412751.8 & $-$300 & 2.7 \\ 
004221.7+412827.6 & $-$300 & 2.8 & 
004226.1+410548.2 & $-$500 & 3.8 & 
004226.4+412811.2 & $-$300 & 3.4 & 
004230.1+412904.0 & $-$300 & 2.8 \\
004230.3+412935.9 & $-$300 & 4.0 & 
004230.9+405714.6 & $-$300 & 3.0 & 
004234.2+413007.3 & $-$300 & 3.0 & 
004235.6+413149.0 & $-$300 & 3.2 \\ 
004238.6+413150.5 & $-$300 & 3.7 & 
004238.9+413135.6 & $-$300 & 2.9 & 
004240.1+410222.7 & $-$300 & 3.5 & 
004241.3+412246.7 & $-$300 & 4.1 \\
004241.9+405155.2 & $-$300 & 3.6 & 
004242.5+410001.4 & $-$300 & 3.0 & 
004242.5+413155.2 & $-$300 & 3.9 & 
004242.9+413159.8 & $-$300 & 3.7 \\
004244.1+413259.2 & $-$300 & 3.4 & 
004244.4+411608.5 & $-$300 & 3.7 & 
004246.2+410111.4 & $-$300 & 2.4 &  
004247.0+413333.0 & $-$300 & 4.0 \\
004247.9+413400.5 & $-$300 & 3.1 &  
004249.1+411945.9 & $-$300 & 3.1 & 
004249.3+412507.5 & $-$300 & 3.2 & 
004251.0+413507.8 & $-$300 & 3.2 \\
{\it 004252.3+410014.8} & $-$300 & 2.1 & 
004252.4+410120.7 & $-$300 & 2.8 & 
{\it 004256.9+413728.1} & $-$300 & 1.6 &
004258.2+410015.9 & $-$300 & 3.1 \\

004300.0+413654.2 & $-$300 & 2.5 & 
004301.5+413717.2 & $-$300 & 2.8 & 
004302.5+413740.5 & $-$300 & 3.5 & 
004306.9+413807.1 & $-$300 & 3.3 \\ 
004311.3+410459.5 & $-$300 & 3.6 & 
004314.2+410033.9 & $-$300 & 2.4 &  
{\it 004324.1+414124.7} & $-$300 & 2.0 & 
{\it 004324.3+414418.7} & $-$300 & 1.9 \\
004324.4+410802.9 & $-$300 & 2.5 &  
004325.6+410206.4 & $-$300 & 4.4 & 
004333.6+411432.3 & $-$300 & 2.8 & 
004334.9+410953.6 & $-$300 & 3.8 \\
{\it 004341.7+414519.4} & $-$300 & 2.0 & 
{\bf 004343.9+411137.6} & $-$300 & 1.8 & 
004346.3+414418.5 & $-$300 & 3.8 &  
004346.8+411239.7 & $-$300 & 2.9 \\ 
004348.1+411133.2 & $-$300 & 3.2 & 
004349.4+411053.8 & $-$300 & 2.4 & 
004354.8+414715.6 & $-$300 & 2.9 & 
004355.8+411211.6 & $-$300 & 4.1 \\
{\it 004356.6+412629.6} & $-$100 & 2.4 & 
004356.8+414831.6 & $-$300 & 3.6 & 
004357.7+414854.0 & $-$300 & 3.1 & 
004358.2+414726.9 & $-$300 & 3.0 \\ 
004358.7+414837.5 & $-$300 & 3.0 & 

004401.5+414909.6 & $-$300 & 3.8 & 
004403.0+414954.7 & $-$300 & 3.1 &  
004403.9+413414.8 & $-$100 & 3.4 \\
004404.9+415016.1 & $-$300 & 2.8 & 
004406.4+412745.0 & $-$100 & 2.9 & 
004407.0+412759.3 & $-$100 & 3.3 & 
004407.8+412110.7 & $-$300 & 3.0 \\
{\bf 004409.5+411856.6} & $-$300 & 1.6 & 
004410.5+420247.5 & $-$100 & 2.9 & 
004412.1+413320.5 & $-$100 & 2.8 & 
004415.3+411905.7 & $-$300 & 2.9 \\
004416.0+414950.7 & $-$100 & 3.1 & 
004418.2+413406.6 & $-$100 & 2.8 & 
004419.2+411930.9 & $-$300 & 2.9 & 
004419.9+412201.2 & $-$300 & 3.2 \\
004420.2+415101.5 & $-$100 & 3.4 & 
004423.0+412050.9 & $-$300 & 3.3 &   
004423.3+413842.6 & $-$100 & 2.8 & 
004423.7+412437.3 & $-$300 & 3.3 \\  
004424.1+412117.3 & $-$300 & 2.9 &  
004424.2+414918.9 & $-$100 & 3.2 & 
{\it 004424.9+413739.1} & $-$100 & 2.1 & 
004426.2+412054.1 & $-$300 & 2.4 \\
004426.7+412729.3 & $-$300 & 2.4 & 
004427.5+413529.8 & $-$100 & 3.0 & 
004429.1+412334.0 & $-$300 & 3.2 &  
{\bf 004430.5+415154.8} & $-$100 & 1.8 \\
004431.1+415110.2 & $-$100 & 3.0 &
004431.9+412233.3 & $-$300 & 2.9 & 
{\it 004431.9+412400.1} & $-$300 & 2.2 &  
{\it 004433.8+415249.7} & $-$100 & 3.5 \\
004435.6+415606.9 & $-$100 & 3.1 & 
004437.9+415154.0 & $-$100 & 3.8 & 
004438.5+412511.1 & $-$300 & 3.7 & 
004443.9+412758.0 & $-$300 & 3.9 \\
004444.1+415359.0 & $-$100 & 3.3 & 
004444.8+412839.9 & $-$300 & 2.6 &  
004448.4+412254.2 & $-$300 & 3.1 & 
004453.0+415340.3 & $-$100 & 3.0 \\
004456.1+412918.2 & $-$300 & 3.2 & 
004456.2+413124.1 & $-$300 & 2.5 &  
004457.2+415524.0 & $-$100 & 3.1 & 
{\it 004458.7+415536.1} & $-$100 & 2.1 \\
004459.1+413233.8 & $-$300 & 4.2 & 
004459.1+414058.5 & $-$100 & 3.1 & 
004459.3+413139.2 & $-$300 & 2.9 & 

004500.7+412836.9 & $-$300 & 4.1 \\
004500.9+413101.1 & $-$300 & 3.0 &  
{\it 004506.1+413615.0} & $-$100 & 2.1 & 
004506.1+415121.0 & $-$100 & 3.4 & 
004506.2+413424.4 & $-$100 & 3.0 \\
004506.9+413407.8 & $-$100 & 2.8 & 
004509.0+415209.7 & $-$100 & 3.3 & 
004511.2+413644.9 & $-$100 & 3.3 & 
004511.3+413633.9 & $-$100 & 2.7 \\
004512.4+413709.6 & $-$100 & 3.2 & 
{\it 004512.8+413531.6} & $-$100 & 2.4 &  
004515.2+413948.5 & $-$100 & 3.1 & 
{\it 004518.5+414013.2} & $-$100 & 2.1 \\
004520.7+414716.7 & $-$100 & 3.1 & 
{\it 004524.4+415537.4} & $-$100 & 2.1 & 
004528.2+414513.6 & $-$100 & 3.1 & 
004536.9+415704.0 & $-$100 & 3.2 \\
004537.2+415802.4 & $-$100 & 2.9 & 
004537.3+415107.0 & $-$100 & 3.2 & 
004537.6+415424.1 & $-$100 & 3.2 & 
004538.5+415231.1 & $-$100 & 3.2 \\
004540.0+415510.2 & $-$100 & 3.2 & 
004541.6+415107.7 & $-$100 & 3.3 & 
{\it 004542.9+415234.8} & $-$100 & 2.5 & 
004543.3+415301.1 & $-$100 & 3.0 \\
004544.3+415207.4 & $-$100 & 3.1 & 

004608.5+421131.0 & $-$100 & 2.9 & 
004617.6+415158.0 & $-$100 & 3.4 & 
004633.6+415932.0 & $-$100 & 3.0 \\
{\it 004634.4+421143.1} & $-$100 & 2.4 &
004641.6+421156.2 & $-$100 & 3.2 & 
& & & & & 
\enddata 
\tablecomments{
206 sources were observed for 5 minutes except for italicized sources (10 min).
Bold sources are \water\ masers; 003918+402158, 
004343+411137, and 004409+411856 were observed
for 15 minutes in two sessions; 004121+404947 was observed for 10 minutes
in a single session; and 004430+415154 was observed for 25 minutes in
three sessions.  No maser variability was apparent.
The rms noise is for 3.3~\kms\/ channels in all cases.
$V_{\rm Obs}$ is the observed central velocity.
% tuning of the 670~\kms\ (50~MHz) bandwidth.
}
\end{deluxetable*}
%}

\section{Observations and Data Reduction}\label{sec:obs}

We observed the $6_{16}-5_{23}$ 22.23508~GHz ortho-water maser line toward
206 24~\mum\/ sources in M31 with the GBT in 2010 October through
December.  A 670~\kms\/ (50~MHz) bandpass was centered on a heliocentric velocity of 
$-300$~\kms\/ for 97 sources in the central parts of the galaxy and
along the minor axis, on 
$-100$~\kms\/ for 52 sources in the redshifted northeast wedge of the galaxy, and on 
$-500$~\kms\/ for 57 sources in the blueshifted southwest wedge of the galaxy (Table \ref{tab:obs}).

Observations were conducted with the dual K-band receivers in a
nodding mode in two circular polarizations with 0.16~\kms\/ (12.2~kHz)
channels and 9-level sampling.    The time on-source was 5 minutes 
except for sources that were re-observed to confirm or refute
possible lines (typically 10 minutes; see Table \ref{tab:obs}).
A winking calibration diode and hourly atmospheric opacity estimates were used for flux
density calibration.  Opacities ranged from 0.05 to 0.13 nepers but
were typically 0.07 nepers.  The estimated uncertainty in the flux
density calibration is $\sim$20$\%$.
Pointing was typically good to within a few arcseconds and the
largest pointing drifts during observations were no more
than 6\arcsec.  
The resolution of the 24~\mum\/ Spitzer image is 6\arcsec\/ \citep{gordon08}, so the unresolved 
IR sources remained within the 33\arcsec\ GBT beam even during the largest pointing drifts.
The 33\arcsec\ beam (FWHM) at 22~GHz spans 125~pc in M31.
%at an assumed M31 distance of 780~kpc.  

After averaging polarizations, 
spectra were Hanning smoothed and subsequently Gaussian smoothed to obtain a final
spectral resolution of 3.3~\kms\ (244~kHz) for the line search.  
Polynomial baselines, typically of fifth order, were fit and subtracted
to obtain flat and generally uniform-noise spectra.
Spectral rms noise measurements for each pointing center are listed in Table
\ref{tab:obs}.
Non-detection spectra generally did not show any features greater than
3$\sigma$, corresponding to roughly 10~mJy, at least an order of magnitude
more sensitive than previous surveys \citep[][Claussen \& Beasley,
priv.\ comm.]{greenhill95,imai01} and likely 
the weakest line that can currently be used for VLBI proper motion 
studies.
All data reduction was
performed in GBTIDL\footnote{GBTIDL (\url{http://gbtidl.nrao.edu/}) is 
the data reduction package produced by NRAO 
and written in the IDL language for the reduction of GBT data.}.
%, and all velocities are given in the heliocentric reference frame.

\section{Results}\label{sec:results}

\begin{figure}
\epsscale{1.15}
\plotone{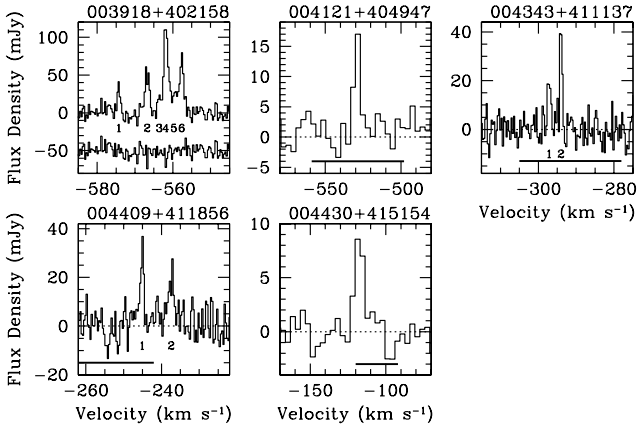}
%\plotone{../plots/all_spectra.ps}
%\plottwo{../plots/He2_10.pdf}{../plots/NGC4038.pdf}
%\plottwo{../plots/NGC4214.pdf}{../plots/NGC5253.pdf}
\caption{Spectra of the five water maser complexes detected in M31.
While all of our spectra were smoothed to 3.3~\kms\ for the search, the lines
of 003918+402158, 004343+411137, and 004409+411856 
are very strong and narrow and we present
their spectra with 0.33~\kms\/ channels.  For 
003918+402158, we plot
fit residuals offset by $-$50~mJy.  Dotted lines on
the other spectra indicate the zero point.
Numbers below the spectra label the lines enumerated in Table
\ref{tab:lines}.  
%Velocities are heliocentric.
The bold horizontal bars indicate the velocity range of CO
within the GBT beam \citep{nieten06}.  
%No CO data are available at the location of 003918+402158, but
%its velocity is roughly consistent with the \ion{H}{1} rotation curve (Figure \ref{fig:rc}).
\label{fig:spectra}}
\end{figure}

We have detected water maser emission in five out of 206
24~\mum-selected regions in M31 (Table \ref{tab:obs}).  
Most of these are clearly maser
complexes that will likely resolve into individual maser spots 
when imaged interferometrically.
Measured line properties are listed in Table \ref{tab:lines}; spectra
are plotted in Figure \ref{fig:spectra}.
The isotropic line luminosities of the individual maser components
are analogous to luminous Galactic water masers.
We also confirm the \citet{sjouwerman10} non-detection of water at the location of the 
methanol maser (Figure \ref{fig:M31map}).

Wherever there is coverage, the water lines correspond to the range of CO (1$-$0)
line velocities within the GBT beam \citep{nieten06}.  The
notable exceptions are 004409+411856, in which the second water maser component
does not overlap with the observed CO line range, and 003918+402158
where there is no CO line coverage.  In the latter,
% bright maser complex --- which lies close to the galaxy's major axis --- 
the observed
maser velocities are broadly consistent with the \ion{H}{1} rotation curve
and its uncertainties 
\citep[][although see \S \ref{sec:discussion-vels}]{chemin09}.

\begin{deluxetable*}{lclrlcl}
\tabletypesize{\scriptsize} 
\tablecaption{Water Maser Line Properties\label{tab:lines}} 
\tablewidth{0pt} \tablehead{
\colhead{Source} & \colhead{Line} & \colhead{$V_{H_2O}$} & 
\colhead{$S_{\rm Peak}$} & \colhead{$\Delta V_{\rm FWHM}$} & 
\colhead{$\int S{\rm d}v$} &
\colhead{$L_{H_2O}$} \\ 
\colhead{} & \colhead{} & \colhead{(km s$^{-1}$)} &
\colhead{(mJy)} & \colhead{(km s$^{-1}$)} & \colhead{(mJy
km s$^{-1}$)} & \colhead{($L_\odot$)} }
\startdata 
003918+402158 
& 1 & $-$574.30(0.07) & 33.5(4.7) & 1.0(0.2) & 37(8) & 0.00052(11) \\
& 2 & $-$566.79(0.04) & 61.7(4.2) & 1.3(0.1) & 87(9) & 0.0012(1)\\
& 3 & $-$563.59(0.08) & 26.9(5.8) & 0.8(0.2) & 23(8) & 0.00032(11)\\
& 4 & $-$561.84(0.02) &122.2(9.1)& 1.13(0.09) & 135(15) & 0.0019(2)\\
& 5 & $-$559.7(0.4)     &  26.0(2.8) & 3.6(1.3) & 99(36) & 0.0014(5)\\
& 6 & $-$557.63(0.04) &  68.5(7.1) & 0.9(0.1) & 66(11) & 0.00092(15)\\
& Sum & & & & 447(43) & 0.0063(6)\\ 
004121+404947
& 1 & $-$528.9(0.4) & 17.7(2.4) & 5.1(0.8) & 95(20) & 0.0013(3)\\
004343+411137
& 1 & $-$297.15(0.10) & 21.1(4.1) & 1.0(0.2) & 23(7) & 0.00032(10)\\ 
& 2 & $-$294.01(0.04) & 44.9(4.8) & 0.74(0.09) & 35(6) & 0.00049(8)\\ 
& Sum & & & & 58(9) & 0.00081(13)\\ 
004409+411856
& 1 & $-$245.09(0.07) & 32.7(4.5) & 1.1(0.2) & 38(8) & 0.00053(11)\\ 
& 2 & $-$237.4(0.2) & 18.1(3.1) & 2.3(0.5) & 44(12) & 0.00062(17)\\ 
& Sum & & & & 82(14) & 0.0011(2)\\ 
004430+415154
& 1 & $-$117.5(0.5) & 9.3(1.4) & 6.8(1.2) & 67(15) & 0.0009(2)
\enddata 
\tablecomments{\
Line properties determined from Gaussian fits.
Isotropic line luminosities were computed from
$L_{H_2O} = (0.023\ L_\odot)\, D^2\int S\,{\rm d}v$, where $D$
is in Mpc and $\int S\,{\rm d}v$ is in Jy km s$^{-1}$ \citep{henkel05}.
Parenthetical values indicate 1$\sigma$ statistical uncertainties.}
\end{deluxetable*}

\begin{figure*}
\epsscale{1.1}
\plotone{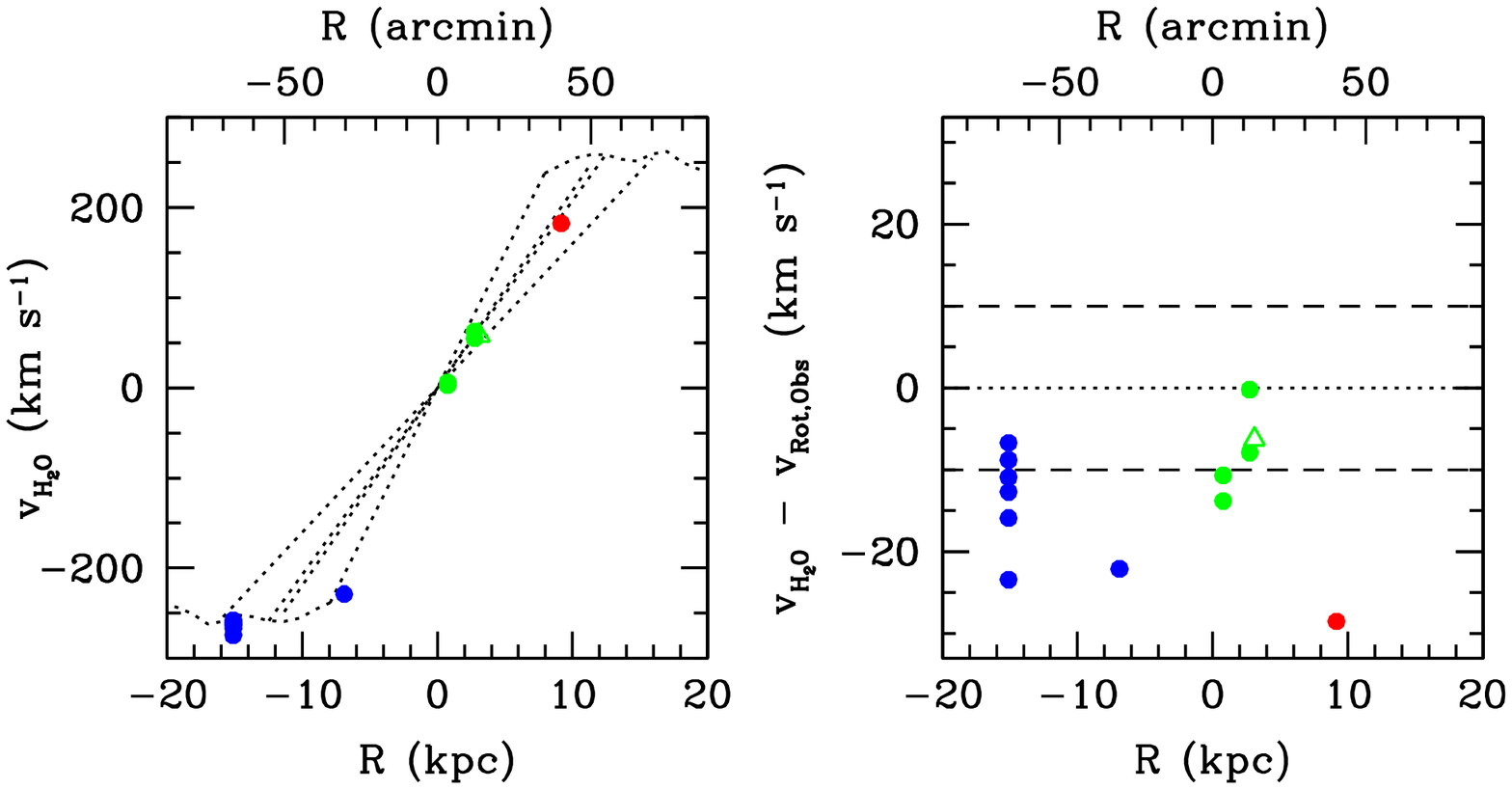}
\caption{Comparison of maser velocities to the \ion{H}{1} rotation curve of
  M31 \citep{chemin09}.
Colors match those in Figure \ref{fig:M31map}, except the methanol maser 
is indicated by a green triangle.
{\it Left:}  Observed water maser velocity versus position along the major 
axis.  The dotted lines indicate the expected observed \ion{H}{1} rotation
curves at the radius of each maser.  Once the radius is reached, the inclined rotation curve is seen 
along the major axis.  
{\it Right:}  The difference between the observed water maser velocities and the projected inclined \ion{H}{1}
rotation velocity versus position along the major axis.  The 
dashed lines indicate a fiducial rotation uncertainty of $\pm$10~\kms, which is 
typical of M31 \ion{H}{1} rotation curve measurements.  The maser 
velocity uncertainties are much less than 1~\kms\ (Table \ref{tab:lines}).
\label{fig:rc}}
\end{figure*}

\vspace{40pt}

\section{Discussion}\label{sec:discussion}

The GBT survey has produced some surprises: (1) the strongest 
water maser complex lies outside of the main star-forming molecular ring in M31, and (2) while 
the 24~\mum\ selection method is a success, the 24~\mum\ luminosity is not predictive of maser presence 
or strength. 
Our survey has revealed no correlation between 24~\mum\ luminosity and 
water maser luminosity, suggesting that while water masers arise in 
star-forming regions, the nonlinear amplification pathways and beamed
nature of the water masers means that they are not predictable based on IR luminosity alone.
We therefore believe that additional strong water masers may be found in M31 
because our survey of 24~\mum-emitting regions has not been exhaustive --- it has simply selected the 
most IR-luminous compact sources (Figure \ref{fig:M31map}).

It is also certain that not all of the selected 24~\mum\ sources are
associated compact \ion{H}{2} regions:  cool stellar atmospheres, supernova
remnants, planetary nebulae, and background AGN
and galaxies can all be sources of 24~\mum\ emission
\citep[e.g.,][]{verley07,mould08}.
Our detection statistics among the observed sample are thus not 
physically meaningful, although the majority of the 24~\mum\ emitters
in the molecular ring of M31 are likely to be compact \ion{H}{2} regions.

\subsection{Detection Rate and Star Formation}

%Water maser variability

%[Comparison to other water masers (Galactic and extragal)]

%{\red [Discuss the ancillary info about the detections:  HI, CO, 24~\mum\/,
%Halpha, radio, etc. --- save a systematic analysis for later?]}

%[maser luminosities...consistent with SF expectations?]

%[typical upper limits on luminosities and consistency with SF]

%Our flux limit (12.5~mJy at 5$\sigma$; see Observations) 
%for this maser search is more than an order of magnitude more sensitive than previous work
%and represents the weakest conceivable line that can currently be used for VLBI proper motion 
%studies.  

%\noindent
%A critical issue is whether we can expect to detect unexceptional
%Galactic analog H$_2$O masers in M31 at the 15~mJy level (5$\sigma$).  
%Detectability is not a problem:  
The sensitivity of the GBT survey corresponds to a $\sim$200~Jy Galactic maser
6~kpc distant.  There are 23 H$_2$O masers in the Palagi \etal\ (1993) sample and
about 30 masers in the \citet{breen10} sample that our survey could
detect at the distance of M31.  Hence, sensitivity is not the only
consideration; the number of water masers detectable
in M31 will also depend on the 
% aggregate population of water masers produced by 
the substantially lower star formation rate of M31 compared to Galactic.
%Given the star formation conditions in M31, how many water masers
%should be detectable at our line flux limit?
%The remaining issue is the number of expected masers 

To estimate the number of expected water masers in Local Group
galaxies, \citet{brunthaler06} used the \citet{greenhill90} Galactic 
water maser luminosity function (LF) scaled from the Galactic star 
formation rate to that of Local Group galaxies, assuming
that maser population is proportional to the star formation rate.  
Local Group water maser
statistics agree fairly well with this approach \citep{brunthaler06}.
We scale the Galactic star formation rate ($\sim$4~$M_\odot$~yr$^{-1}$; Diehl \etal\ 2006) 
to that of M31 ($\sim$0.35~$M_\odot$~yr$^{-1}$; Walterbos \& Braun
1994, Tabatabaei \& Berkhuijsen 2010) 
to predict 2.3 water masers for an isotropic line luminosity limit of 
% $7\times10^{-4}$~$L_\odot$,  %(7.4x10^-4 to be precise) 
$4\times10^{-4}$~$L_\odot$,  %(4.4x10^-4 to be precise) 
which is a 3$\sigma$ line flux limit of 9~mJy and width 3.3~\kms.
This small number suffers from small number statistics, but 
our five detections suggest that either we have detected 
the majority of bright water masers in M31 or that the star formation 
scaling approach is inaccurate.
This program is the first to detect the tail of the
H$_2$O maser LF in M31.

\subsection{Velocities}\label{sec:discussion-vels}

A comparison of the maser velocities to the rotation of M31 is of
special interest for a geometric distance determination.  Proper rotation 
is a measure of angular frequency, and if this
angular motion corresponds directly to the rotation velocity, then 
a geometric distance may be obtained:  $D = V_{Rot}/\mu_{Rot}$.
%The issue of paramount importance for a 
A proper rotation distance
determination relies on the the masers rotating with the galaxy, 
% which is likely, 
but they could also exhibit their own peculiar proper motions,
so a large ensemble of water masers would be ideal to 
reduce this source of error.  

In Figure \ref{fig:rc} we examine the correspondence between the maser
velocities and the projected and inclined \ion{H}{1} rotation curve
\citep{chemin09}.  
Since the maser velocities are extremely well-determined
(sub-\kms), all of the uncertainty in the velocity comparison arises 
from the innate velocity dispersion of the disk, the 
statistical uncertainty in the rotation curve ($\sim$10~\kms), and 
the systematic uncertainties such as inclination, which can 
induce a rotation curve offset of order 10--15~\kms\ \citep{corbelli10}.
While the masers show fairly good agreement with the \ion{H}{1} rotation,
there is a notable overall blueshift that may be indicative of preferentially 
oriented maser amplification pathways associated with outflows.  
But given the propensity for masers to show peculiar 
velocities, a larger sample is needed to rigorously examine this effect.

%For the rotation curve we assume 
%azimuthal symmetry and constant inclination, which is probably 
%incorrect given the disturbed nature of the molecular ``ring'' in M31.

%Figure \ref{fig:rc} shows a trend in observed velocity-rotation curve offset
%from the receding to the approaching side of the
%disk suggestive of a non-azimuthally symmetric rotation curve or
%a warped disk.  But given the propensity for masers to show peculiar 
%velocities, a larger sample is needed to rigorously examine these possibilities.

\subsection{Proper Motions}

%Now that water masers have at last been detected in the Andromeda Galaxy, it is time to 
%completely survey the galaxy for masers, to localize and map the maser complexes, and to begin 
%precision astrometry observations in order to measure proper motions of maser spots, the three- 
%dimensional velocity of M31, and its geometric distance.
%Only one maser is required to obtain a coarse proper motion measurement, 
%but more than one would allow us to obtain a precise proper motion as
%well as the rotation of M31 (and hence a geometric distance, following 
%the work of Brunthaler \etal\ 2005 on M33).
 
%Moving cluster method for distance?  No (but mention)

%Geometric distance from rotation?  Yes.

%Expectations for various proper motions.

%What contributes to the proper motion of masers in M31, and what information do
%proper motion measurements provide?  
The three main contributions
to the proper motion of masers in M31 are (1) the rotation of the spiral
disk, (2) the systemic proper motion of M31 itself, and (3) the relative motion
of masers omitting the galaxy's rotation:

(1)  The direct observation of the rotation of the M31 spiral disk 
is the (in)famous van Maanen experiment
\citep{vanmaanen23,vanmaanen35}.  A rotating disk seen edge-on will 
show no proper motion at the tangent points and the full circular
velocity along the minor axis.  M31's inclined disk 
has a circular velocity at the molecular ring radius of
about 250~\kms, which becomes a transverse motion of 56~\kms\/
at the tangent points or 15 microarcseconds ($\mu$as) year$^{-1}$.  
The transverse motion along the minor axis is expected to be about 
70~\muasyr.  Since the rotation curve of M31 is well-known
\citep[e.g.,][]{chemin09,corbelli10}, the 
measurement of the proper motions of masers distributed
around the disk will provide a geometric distance to the galaxy,
limited roughly equally by the uncertainty in the rotation model
and in the proper motion measurements.

(2)  The transverse motion of M31 provides the two unknown
components of its three-dimensional velocity.  % with respect to the Galaxy. 
% (or the cosmic microwave background).    
%The measured quantity is the proper motion with respect to our local reference
% frame, which can be corrected to the center of the Galaxy \citep[e.g.,][]{brunthaler05}.
The expected transverse velocity of M31 is
of order 100~\kms\/ \citep[][but see van der Marel \& Guhathakurta 2008]{loeb05,brunthaler07},
or 27~\muasyr, and 
measurable over two or three years (see below).  
%The transverse velocity of M31 is
%the keystone to Local Group dynamics and dark matter.

(3)  There may be relative motions of masers within M31 
due to peculiar deviations from the bulk rotation of the disk and due to 
the line of sight motion of M31 itself.  The former peculiar motions
are of order 10--20~\kms\ (Figure \ref{fig:rc}), which is significantly smaller
than the velocities discussed above and of the same order
as the uncertainty in the rotation curve.  These will tend to average 
out over an ensemble of maser complexes.  The latter is 
the moving cluster effect caused by M31's radial velocity toward the
Galaxy;  M31 gets areally larger as it nears, causing
relative maser separations to grow over time.  This effect is
small, about 3~\muasyr\ of divergence for the maser pair with the largest 
separation, 1.79$^\circ$.  The size of this effect is equivalent to $\sim$10~\kms, 
similar to maser peculiar motions.
It is possible that with long time baselines and a rich network
of masers one could extract the distinct signal of diverging maser
spots and thus estimate the distance to M31 (but with lower precision
than the proper rotation method).

% These are typically 8--10~kpc distant and
% have flux densities greater than 100~Jy {\red [fix stats]}.  
% For example, the exceptional $>$1000~Jy H$_2$O 
% maser in W49 N (11~kpc distant; Gwinn, Moran, \& Reid 1992) would be at least
% a 200~mJy maser in M31.  

% xxx xxxx (23 Galactic above our
% luminosity limit in Palagi et al 1993; Breen etal 2010 have about 27(+3) detectable masers
% at M31, although distance remains ambiguous, we assume a conservative typical value of 6 kpc, 
% which translates our sensitivity into about 200 Jy; extant water maser surveys are consistent
% with the LF predictions for numbers of detections given their biases and incomplete coverage
% of the Galaxy --- the largest uncertainty in any prediction method lies in water maser 
% variability (Breen \etal\ 2010)); 
% the remaining issue is the number of expected masers detectable at our line flux limit.  

How long will it take to
obtain proper motions, and what are the limits on the proper motions
that can be measured?  The \citet{brunthaler05} study of the proper motion of
water masers in M33 informs our expectations for M31.  
Water masers appear in many-component complexes that can show 
dramatic variability, including the appearance of new masers
and the disappearance of others.  But over the span of about three
years, \citet{brunthaler05} were able to track several persistent
maser features and obtain proper motions relative to 
background compact radio sources.
% that are radially distant but nearby on the sky (providing a 
%fixed astrometric reference frame).  
Their astrometric
uncertainties were on average 8--10~$\mu$as
in each epoch. %of 20--24 hour integrations.  
Proper motion %measurement
uncertainties from a three-year fit were $\sim$5~\muasyr,  
which would correspond to a transverse velocity of 18~\kms\/ in M31.  

%Our planned VLBI observations will benefit from two recent
%developments:  22 GHz receiver upgrades at the VLBA and EVLA,
%improving system temperature by nearly a factor of 2, and 
%the inclusion of the phased EVLA and the GBT with the VLBA to form the HSA, 
%further increasing sensitivity by roughly a factor of 4.  
We estimate
that in two 12-hour observations with the High Sensitivity Array one can obtain 1~mJy
rms noise in 1~\kms\/ channels.
% (two sessions spaced by roughly a week will provide an assessment of errors unrelated to 
%proper motion).  
This translates into a 6~\muas\/ 
uncertainty {\it per epoch} for a 40~mJy line.
% (three such complexes have been detected; Figure \ref{fig:spectra}).
%, but only two are needed to obtain a geometric distance to M31).  
%Using the proper motion estimates 
%delineated above, 
In a single year we expect 
% this precision would provide 
a 10$\sigma$ detection of the proper rotation
along the minor axis, a 2.5$\sigma$ detection
along the major axis, and a 4.5$\sigma$ detection of the
systemic transverse proper motion of M31 (assuming 100~\kms).
Note that proper motion measurement errors improve
as $t^{3/2}$ \citep{reid09}, so a 2--3 year astrometry campaign should be 
able to determine with high significance both the geometric distance
to M31 and its transverse velocity.

\section{Conclusions}

We have identified the first H$_2$O masers in the M31
based on pointed observations of 24~\mum-selected regions.
The 24~\mum\/ selection method clearly succeeds in identifying 
water masers, but the 24~\mum\/ properties are not predictive of maser
presence or brightness.  We suggest that the catalog of water masers
in M31 is incomplete and that an exhaustive survey of IR-luminous regions
will identify additional masers.
%consistent with the aggregate star formation rate of M31.

The newly identified maser complexes are suitable
for proper motion studies of M31.  The next steps for proper motion 
and geometric distance measurements are interferometric 
localization and mapping followed by VLBI mapping and 
monitoring.  We predict that proper rotation and systemic proper
motion will be measurable in a few years with current facilities, and
that the ``moving cluster'' divergence of maser spots as M31
approaches the Galaxy may be detectable in the long term.  

%Ancillary uses for the masers detected in M31 include general maser studies, mapping
%star-forming regions, and the direct detection of in-situ magnetic fields via Zeeman splitting.
%M31 has a rich library of multispectral observations that can be compared to maser-active regions
%for studies of star formation, molecular gas excitation, and dust
%extinction.  [keep this?]

\acknowledgements
The author thanks K.\ Gordon for the Spitzer map,
M.\ Claussen and T.\ Beasley for sharing their results, 
K.\ Willett for proposal feedback, and the anonymous referee.
This research has made use of the NASA/IPAC Extragalactic Database
(NED) and uses observations made with the {\it Spitzer Space Telescope},
both of which are operated by the Jet Propulsion Laboratory,
California Institute of Technology, under a contract with
NASA. 

{\it Facilities:} \facility{GBT ()}

%\clearpage

\end{document}